\documentclass[intlimits,twoside,a4paper]{article}

\usepackage[cp1251]{inputenc}
\usepackage[eqsecnum]{cmpj3}

\usepackage{bm}

\issue{2021}{24}{1}{13701}
\doinumber{10.5488/CMP.24.13701}

\title[Electronic properties in bilayer graphene-like]%
{Electronic properties of bilayer sheets forming moir\'e patterns%
}
\author[W.S. Wu-Mei, R.R. Rey-Gonz\'alez]{W.S.
        Wu-Mei\footnote{wswum@unal.edu.co},  R.R. Rey-Gonz\'alez\footnote{rrreyg@unal.edu.co}}
\address{National university of Colombia, Bogota, Colombia}

\date{Received December 12, 2020, in final form August 26, 2020}
\begin{document}

\maketitle

\begin{abstract}
In this article, we report the electronic band structures of hexagonal bilayer systems, specifically, rotated graphene-graphene and boron nitride-boron nitride bilayers, by introducing an angle between the layers and forming new periodic structures, known as moir\'e patterns. Using a semi-empirical tight-binding approach with a parametrized hopping parameter between the layers, using one orbital per-site approximation, and taking into account nearest-neighbor interactions only, we found the electronic dispersion relations to be around $K$ points in a low energy approximation.
Our results show that graphene bilayers exhibit zero band gap for all angles tested in this work. In boron nitride bilayers, the results reveal a tunable bandgap that satisfies the prediction of the bandgap found in one-dimensional diatomic systems presented in the literature.

\end{abstract}

By exploiting the possibility of obtaining different conducting properties of two-dimensional bilayer systems by rotating one of the layers, it is possible to generate new periodicities, also known as moir\'e patterns \cite{Groups, Beyer_stm}. In this work, we study hexagonal bilayer systems, mainly two types of bilayers. The first one is composed of graphene-graphene layers, the second, boron nitride-boron nitride layers. As the theory predicts, the electronic band structure of an arrangement of atoms might change as the periodicities pattern changes \cite{Ibach}. For instance, we can get different bilayer patterns by introducing different angles between layers \cite{Walleca, Kittel}, forming new periodicities. However, given the size and complexity of the new systems formed, we must take into account the size of a lattice, compared with the angle between layers. The experimental relation between layers $\theta$ and the size of the pattern $L$ is known as a moir\'e pattern, mentioned in 1998 by H. Beyer et al. \cite{Beyer_stm}, who experimentally measured the lattice constant $L$ of a moir\'e pattern, generated in a graphite bulk [LHS\eqref{eq_Moire}], 
\begin{equation}\label{eq_Moire}
  L(\theta)=\begin{cases}\frac{1}{2}
 \frac{a}{ \sin(\theta/2)}\hspace{10mm} 0<\theta\leqslant \piup/3,\\
a \hspace{25mm}\theta=0.
  \end{cases} \hspace{10mm}
  N=\begin{cases}
\frac{1}{\sin^2\left(\theta /2\right)}\hspace{10mm}0<\theta\leqslant \piup/3,
\\ \hspace{5mm}1 \hspace{20mm}\theta=0.
\end{cases}
\end{equation}
Later, S. Shallcross et al. theoretically studied the electronic properties of the graphene bilayers for a given set of angles \cite{Shallcross_turbostratic, Shallcross_intereference}. They proposed a theorem that relates the periodicity formed with the angle $\theta$ between layers \cite{Shallcross_twist}. The commensuration theorem is condensed in [RHS \eqref{eq_Moire}]. Using this theorem, we calculate a set of angles and lattice sizes presented in table \ref{tabla_triangular}.
 \begin{table}[!htb]
 \caption{The number of atoms $N$ per unit cell with graphene bilayer lattice constant $L_{g}$ and nitride boron bilayer lattice constant $L_{nb}$.}\label{tabla_triangular}
  \centering
\begin{tabular}{|l|l|l|l|ll|}
\cline{1-4}
 $\theta$ & $N$   & $L_{g}$(\AA) & $L_{nb}$(\AA) & \\ \cline{1-4}
 60.0              & 4   & 2.46 & 2.51 &  \\ \cline{1-4}
 21.79            & 28  & 6.51 & 6.65 &  \\ \cline{1-4}
 13.17         & 76  & 10.72 & 10.95 &  \\ \cline{1-4}
 9.43         & 148 & 14.96 & 15.28 &  \\ \cline{1-4}
 0.0             & 4   & 2.46& 2.51 &  \\ \cline{1-4}
\end{tabular}
\end{table}
Other studies include Shallcross et al. \cite{Shallcross_turbostratic}, whose article only studied graphene-graphene band structures. R. M. Ribeiro et al. \cite{Ribeiro_stability_BN} calculated boron nitride band structures for a finite size bilayer using DFT methods. Guohong Li and J. M. B. Lopes dos Santos \cite{Guohong_Hove, Lopez_dos_santos_continiuum} used an effective Hamiltonian method to find the electronic properties of bilayer systems for angles ($\theta\leqslant 10$). However, given the complexity of the systems and the computational resources required to get this job done, they calculated band structures for a limited number of atoms. In this work, we computed band structures of bilayer systems by using semi-empirical tight-binding approximations with low energy approximations around $K$ points \cite{Rafi_double}, using one orbital per site and counting only the first neighbor approximation.

\section*{Mathematical model employed}

We modelled the system as a perfectly flat and indefinitely extended bilayer. Hence, we neglected all sorts of thermodynamic fluctuations and border effects. The Hamiltonian of the system can be written as the sum of two isolated, tight-binding monolayers, $H_1$ and $H_2$, interacting via Van der Waals interaction $V$ as follows:
\begin{eqnarray}
H=H_1+H_2+V&=&-\gamma_{0}\sum_{i,j}(a_{mi}^\dagger b_{m,j}+H.c.)-\gamma_f\sum_{i,j}(a_{1,i}^\dagger a_{2,j}+H.c.)\\
\nonumber
&-&\gamma_4\sum_{i,j}(a_{1,i}^\dagger b_{2,j}+a_{2,i}^\dagger b_{1,j}+H.c.).
\label{eq_Hamilton}
\end{eqnarray}
The parameter $\gamma_f$ we had introduced here, depends on the relative distance between a pair of points~$i,j$, and in the type of interaction between each site, located in a sub-lattice $\textbf{A}_{m',i}$ or $\textbf{B}_{m,j}$ of a layer $m$. On the contrary, a type $\gamma_0$ is the hopping parameter between an electron located on a site $\textbf{A}_{1,i}$ to a site $\textbf{B}_{1,j}$ of the same layer, $\gamma_3$ is the hopping parameter between the site $\textbf{A}_{1,j}$\textendash $\textbf{A}_{2,j}$ from different layers, and the $\gamma_4$ corresponds to the hopping parameter between the sites $\textbf{A}_{1,i}$\textendash $\textbf{B}_{2,j}$ from the two different layers as shown in Castro Neto et al., \cite{Castro_neto_electronic}, 
where all hopping parameters are fixed, and the relation dispersion was
calculated for two types of possible setup, hence for $\theta=0^\circ$ and $\theta = 60^\circ$. In this work, we extended this definition into a new  hopping parameter $\gamma_f(\bf{r})$  that depends on the relative distance $\bf{r}$ between sites. This hopping parameter also fits into the values of $\gamma_3$ and $\gamma_4$ when it is evaluated at $(0,a)$ and $(a,a)$. Hence, 
$\gamma_f(0,a)=\gamma_3$ and $\gamma_f(a,a)=\gamma_4$
\begin{equation}
\gamma_f(\bf{r})=\frac{\gamma_3-\gamma_4}{a}\sqrt{(x)^2+(y-a)^2}+\gamma_4.
\end{equation}
The lattice constant for each system is $a$, while $x$ and $y$ are the relative distances between the two sites. This contrasts with Castro Neto et al. \cite{Castro_neto_electronic}. In the equation~(\ref{eq_Hamilton}), the operators $a_{m,i}$ and $ b_{m,i}$ are defined as the charge carrier annihilation operators in the sublattices $\textbf{A}_{m',i}$ and $ \textbf{B}_{m,j}$, respectively. Meanwhile, the operators $a_{m,i}^\dagger $ and $ b_{m,i}^\dagger$ are defined as the construction operators. $\gamma_0$ is defined as the in-layer hopping parameter.
The hopping parameters for graphene-graphene and boron nitride-boron nitride bilayers  \cite{Castro_neto_electronic,Ribeiro_stability_BN} are resumed in the table \ref{tabla_resumen}. For boron-nitride bilayers, we only consider $B-N$ interactions. Hence, $N-N$ and $B-B$ hopping parameters in between layers are set arbitrarily to zero \cite{Ribeiro_stability_BN}.
\begin{table}[!t]
\caption{List of constants used in this paper. Hopping parameters $\gamma_0$, $\gamma_1$, $\gamma_3$, and $\gamma_4$~\cite{KawaguchiBC6N,Ribeiro_stability_BN}. Distance between monolayers $d$ and inter-atomic distances $a_0$, for graphene and nitride boron bilayer~\cite{Dresselhaus_Intercalation}.}
\centering
\vspace{4mm}
\begin{tabular}{|c|c|c|cc|}
\cline{1-3}
        &  Graphene  & Boron nitride &  \\ \cline{1-3}
$\gamma_0$ & $2.8$~eV    & $2.33$~eV &  \\ \cline{1-3}
$\gamma_1$ &  $0.4$~eV    &  $ 0.32$~eV &  \\ \cline{1-3}
$\gamma_3$ & $0.3$~eV        &  $0.25$~eV &  \\ \cline{1-3}
$\gamma_4$ & $0.04$~eV   & $N/A$ &  \\ \cline{1-3}
$d$& $ 3.35$\AA & $ 3.75$\AA&\\ \cline{1-3}
$a_0$& $ 1.42$\AA & $ 1.45$\AA&\\ \cline{1-3}
\end{tabular}\label{tabla_resumen}
\end{table}

\begin{figure}[!t]
  \centering
  \includegraphics[scale=0.2]{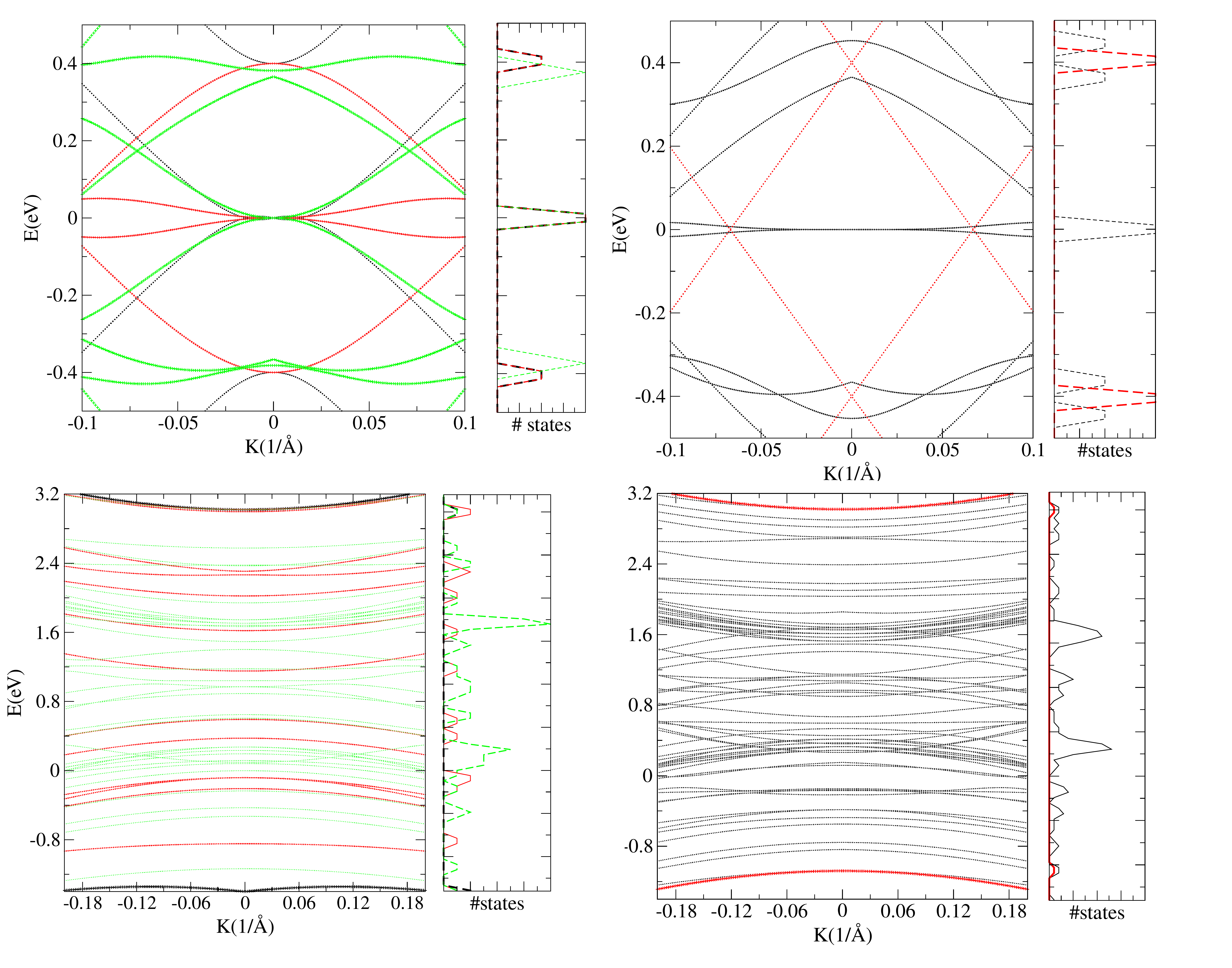}
  \caption{(Colour online) Low energy approximation of electronic band structure around Dirac points $K=(2\piup/3a, 2\piup / 3\sqrt{3}a)$, with its respective projected DOS evaluated in K. Upper left: Graphene-graphene bilayer black $60^\circ$, red $21.79^\circ$ and green $13.17^\circ$. Upper right: Graphene-graphene bilayer black $9.43^\circ$ and red $0.0^\circ$. Down left: Nitride boron-nitride boron bilayer black $60^\circ$, red $21.79^\circ$ and green $13.17^\circ$. Down right:  Nitride boron-nitride boron bilayer black $9.43^\circ$ and red $0.0^\circ$.} \label{im_rec_hex}
\end{figure}

\begin{figure}[!t]
  \centering
  \includegraphics[scale=0.28]{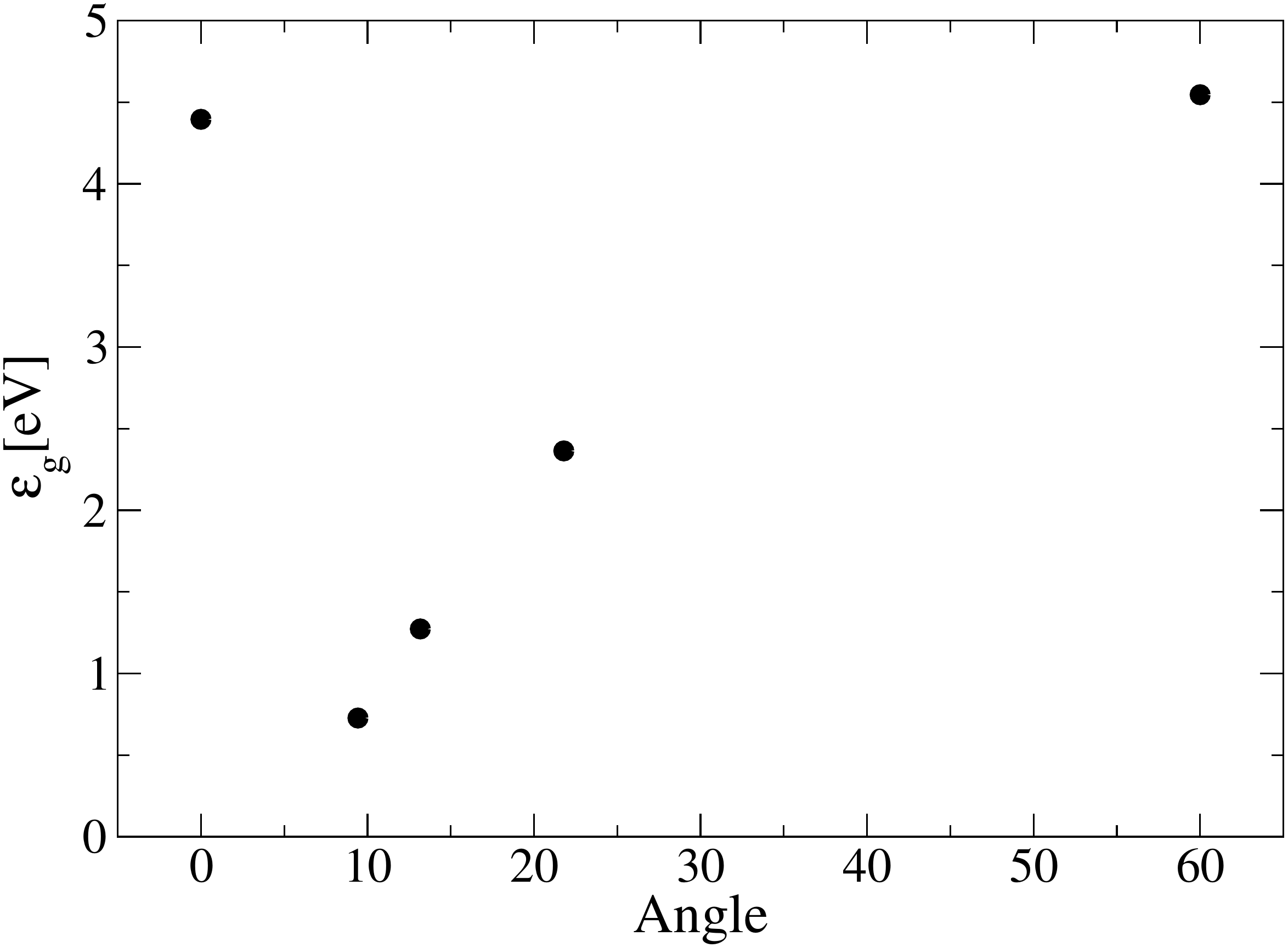}
  \caption{Bandgap behavior of nitride boron bilayer.}
  \label{fig_gap}
\end{figure}

We can write the Hamiltonian \eqref{eq_Hamilton} as a $N\times N$ matrix for each system. Solving the eigenvalues of this equation, we find the band structure for each bilayer, and angle around Dirac points \ref{im_rec_hex} resumed in the graphics \ref{fig_gap}.

As shown in the graphic \ref{im_rec_hex} and compared with the existing literature \cite{Castro_neto_electronic}, the structure of the graphene bilayer for $\theta=0$ holds a characteristic two shifted Dirac cone, matching the theory. However, for $\theta=60$, it shows parabolic behavior with a nonnull charge carrier effective mass around the Dirac point $K$. Moreover, the following matches the previous theoretical prediction \cite{Castro_neto_electronic,Dos_santos_twist_electronic_structure}, confirming the validity of the methods employed here in the two limits. However, in the intermediate values between $60$ and $0$ degrees, it reveals degenerated bands.

On the other hand, the electronic band structure of boron nitride shows that it is possible to modulate the bandgap behavior by changing the angle between layers, as demonstrated in figure \ref{fig_gap}.

\section*{Conclusions}
In conclusion, we have shown that it is possible to tune the bilayer bandgap by changing specific geometrical settings. 
This result is only possible if the atoms of each layer are made up of two
different species of atoms.
The new geometrical pattern obtained after rotating each layer must be different from the previous one. Considering that we formerly assumed perfectly flat layers and boundaryless bilayers, the predictions are ideally valid at the center of relatively large samples stored at low temperatures.
 
This might be a feasible way to create bilayered semiconductors, allowing us to modify the bandgap values by modifying their relative angles. In this case, the boron nitride--boron nitride bilayer displays a bandgap modulation within the range of $0.73$ eV to $4.5$ eV, considering that we only took into account $N-B$ type interactions between bilayers. We encourage measurements on these systems through this work, which will allow us to confirm or refute our predictions by recreating boron nitride bilayers and graphene bilayers in low-temperature setups.

 \section*{Data availability}
 The C++ codes used to compute the bands structures of this study have been deposited in the GitHub repository (https://github.com/wswum3009/Electronic-properties-of-bilayer-sheets-forming-Moire-pat\-terns).


\ukrainianpart
\title{Електронні властивості двошарових листів, що формують муарові малюнки%
}
\author{В.С. Ву-Меї, Р.Р. Рей-Гонзалес}
\address{Колумбійський Національний університет, Колумбія}

\makeukrtitle 

\begin{abstract}
У цій статті описано електронні зонні структури гексагональних двошарових систем, зокрема, обертових двошарів графен-графен та ``нітрид бору-нітрид бору'', вводячи кут між шарами та формуючи нові періодичні структури, відомі як муарові малюнки. Використовуючи напівемпіричний підхід сильного зв’язку з параметризованим  параметром перескоку між шарами та одноорбітальне одновузлове наближення і, враховуючи лише взаємодії найближчих сусідів, встановлено, що співвідношення електронної дисперсії знаходяться на рівні $K$ точок у наближенні  низької енергії.
Наші результати показують, що двошари графену мають нульову заборонену зону для всіх кутів, протестованих у цій роботі. У двошарах нітриду бору  виявлено налаштовувану заборонену  зону, що підтверджує передбачення  для забороненої зони, виявленої в одновимірних двоатомних системах, представлених у літературі. 	
\keywords наближення сильного зв’язку, графен, нітрид бору, двошар, муарові малюнки, вандерваальсівські взаємодії, теорема сумірності, наближення низької енергії
\end{abstract}

 \lastpage

\end{document}